\documentclass[12pt]{iopart}
\usepackage{iopams}
\usepackage{amsthm}
 \usepackage[vcentermath]{youngtab}
\eqnobysec 

\theoremstyle{plain}
\newtheorem{thm}{Theorem}[subsection]
\newtheorem*{thm*}{Theorem}
\newtheorem*{G-R}{Gale--Ryser Theorem}

\newtheorem*{claim*}{Claim}
\newtheorem{cor}[thm]{Corollary}

\newtheorem{prob}[thm]{Problem}

\theoremstyle{definition}

\swapnumbers
\theoremstyle{remark}
\newtheorem{rem}[thm]{Remark}
\newtheorem{Ex}[thm]{Example}
\newtheorem*{Ex*}{Example}

\newtheorem*{rem*}{Remark}

\begin{document}

\title[Quantum marginal problem]{Quantum marginal problem and N-representability}

\author{Alexander A. Klyachko\dag }

\address{\dag\ Department of Mathematics, Bilkent University,
Bilkent, Ankara, 06800 Turkey}

\ead{klyachko@fen.bilkent.edu.tr}

\begin{abstract}
A variant of the quantum marginal problem
was known
from early sixties as {\it
$N$-representability problem}. In 1995 it was designated by
National Research Council of USA as one of ten most prominent
research challenges in quantum chemistry.
In spite of this recognition the progress was very slow, until  a
couple of years ago the problem came into focus again, now in
framework of quantum information theory. In the paper I give an
account of the recent development.
\end{abstract}


\maketitle


\section{Introduction}
\texttt{}The quantum marginal problem is about relations between
spectrum of mixed state $\rho_{AB}$ of two (or multi) component
system $\mathcal{H}_{AB}=\mathcal{H}_A\otimes
\mathcal{H}_B$ and that of reduced states $\rho_A$ and $\rho_B$ of
the components. This problem emerges a couple of years ago in
framework of quantum information theory.

The problem can be stated in plain language as follows. Let
$M=[m_{ijk}]$ be complex cubic matrix and $M_1,M_2,M_3$ be Gram
matrices formed by Hermitian dot products of parallel slices of
$M$. We are seeking for
relations between spectra of these matrices.

In section 3 we survey some   recent results that laid the ground
of the quantum marginal problem and discuss a solution of this
problem based on geometric invariant theory.

A variant of
QMP dealing with system of $N$ fermions,
like electrons in an atom or a molecule, was known from early
sixties as $N$-representability problem (Coulson, 1960). Its
solution allows to calculate nearly all properties of matter which
are of interest for chemists and physicists (Coleman and Youkalow
2000). By this reason it was designated as one of ten most
prominent research challenges in theoretical chemistry (Stillinger
\etal 1995). In section 3 we outline a general solution of the
$N$-representability problem for one particle reduced density
matrix. For systems of rank $\le 8$
explicit sets of linear constraints  are given for pure state $N$
representability. A representation theoretical interpretation of
$N$-representability plays crucial role in the calculations.

The
reduction
$\rho_{AB}\mapsto\rho_A$ is known in mathematics as contraction.
For example, Ricci curvature of a Riemann manifold is a
contraction of its Rienmann curvature. The results of this paper
imply   some
inequalities
between spectra of Riemann and Ricci curvatures, see Example
\ref{Riem_Ric1}. Recall that in general relativity Ricci curvature
is governed by the energy-momentum tensor, i.e.  by physical
content of the space, while Riemann curvature is responsible for
its geometry and topology. The above constraints impose some
bounds on influence of matter on geometry.


\section{Classical marginal problem}\label{Bell}
\subsection{Marginal disributions}
Let's start with {\it classical marginal problem\,} which  asks
for existence of a ``body" in $\mathbb{R}^n$ with given
projections onto some coordinate subspaces
$\mathbb{R}^I\subset\mathbb{R}^n, I\subset
\{1,2,\ldots,n\}$, i.e. existence of {\it probability density}
$p(x)=p(x_1,x_2,\ldots,x_n)$ with given {\it marginal
distributions}
\begin{eqnarray*}p_I(x_I)=\int_{\mathbb{R}^J}p(x)dx_J, J=\{1,2,\ldots,n\}\backslash I.
\end{eqnarray*}

In discrete version the classical MP amounts to calculation of an
image of a multidimensional symplex, say
$\Delta=\{p_{ijk}\ge0|\sum p_{ijk}=1\}$, under a linear map like
\begin{eqnarray*}\pi:\quad \mathbb{R}^{\ell m
n}&\rightarrow&\mathbb{R}^{\ell
m}\oplus\mathbb{R}^{ mn}\oplus\mathbb{R}^{n\ell},\\
\qquad p_{ijk}&\mapsto&(p_{ij},p_{jk},p_{ki}),\\
p_{ij}=\sum_k &p_{ijk}&,\quad p_{jk}=\sum_i p_{ijk},\quad
p_{ki}=\sum_j p_{ijk}.
\end{eqnarray*}
The image $\pi(\Delta)$ is a convex hull of
the projections of vertices of $\Delta$.  So the classical MP
amounts to calculation of facets of a convex hull. In high
dimensions it might be a computational nightmare (Pitowsky 1989).

\subsection{Classical realism}
\label{class_real}
Let $X_i:\mathcal{H}_A\rightarrow \mathcal{H}_A$ be observables of
quantum system $A$. Actual measurement of $X_i$ produces random
quantity $x_i$ with values in ${\rm Spec\,}(X_i)$ and density
$p_i(x_i)$ implicitly determined by expectations
\begin{eqnarray*}\langle f(x_i)\rangle=\langle\psi|f(X_i)|\psi\rangle
\end{eqnarray*}
for all functions $f$ on spectrum ${\rm Spec\,}(X_i)$.
 For
{\it commuting} observables $X_i,i\in I$ the random variables
$x_i, i\in I$ have {\it joint distribution} $p_I(x_I)$ defined by
similar equation
\begin{equation}\langle
f(x_I)\rangle=\langle\psi|f(X_I)|\psi\rangle,\quad \forall
f.\label{joint}
\end{equation}
{\it Classical realism} postulates existence of a hidden joint
distribution of {\it all variables} $x_i$. This amounts to
compatibility of the marginal distributions (\ref{joint})
for {\it commuting} sets of observables
$X_I$. 
{\it Bell inequalities}, designed to test classical realism,
stem from the classical marginal problem.

\begin{Ex}
Observations of disjoint components of composite system
$\mathcal{H}_A\otimes\mathcal{H}_B$ always commute. For two qubits
with two measurements per site their compatibility
is given by {16} inequalities obtained from {\it
Clauser-Horne-Shimony-Holt} inequality (Calauser {\etal} 1969)
\begin{eqnarray*}\langle a_1b_1\rangle+\langle a_2b_1\rangle+\langle
a_2b_2\rangle-\langle a_1b_2\rangle+2\ge 0
\end{eqnarray*} by spin flips
$a_i\mapsto \pm a_j$ and permutation of the components
$A\leftrightarrow B$. Here $\langle a_ib_j\rangle$ is expectation
of product of spin projections onto directions $i,j$ in sites
$A,B$.

\end{Ex}
\begin{Ex}   For three qubits with two
measurements per site the marginal constraints amounts to {53856}
independent inequalities (Pitowsky {\etal} 2001).
This example may help to
disabuse us from overoptimistic expectations for the {\it quantum
marginal problem} to be discussed below.
\end{Ex}
\begin{Ex}\label{Gale} Univariant marginal distributions are always compatible, e.g. we can
consider $x_i$ as independent variables. However under additional
constraints, say
for a ``body" of constant density, even univariant marginal
problem becomes nontrivial. For its disctete version
{\it
Gale-Ryser theorem}
(Gale 1957) tells that partitions $\lambda,\mu$ are margins of a
rectangular $0/1$ matrix iff majorization inequality
$\lambda\prec\mu^t$ holds. Here marginal values
arranged in decreasing order are treated as {\it Young diagrams\,}
{
$$\lambda=(5,4,2,1)=\yng(5,4,2,1)\quad\lambda^t=(4,3,2,2,1)=\yng(4,3,2,2,1)$$}
$\mu^t$ stands for {\it transposed\,} diagram,  and the {\em
majorization order\,} $\lambda\prec\nu$ is defined by inequalities
$$
{\begin{array}{rcl}\lambda_1&\le&\nu_1\\
\lambda_1+\lambda_2&\le&\nu_1+\nu_2\\
\lambda_1+\lambda_2+\lambda_3&\le&\nu_1+\nu_2+\nu_3\\
\cdots\cdots\cdots\cdots\cdots&\cdots&\cdots\cdots\cdots\cdots\cdots
\end{array}}
$$
\end{Ex}

\section{Quantum marginal problem}
\subsection{Reduced states}
Density matrix of composite system $AB$ can be written as a linear
combination of separable states
\begin{equation}\label{sep_decomp}
\rho_{AB}=\sum_\alpha a_\alpha \rho_A^\alpha\otimes \rho_B^\alpha,
\end{equation}
where $\rho_A^\alpha, \rho_B^\alpha$ are mixed states of the
components $A,B$ respectively, and the coefficients $a_\alpha$ are
{\it not\,} necessarily positive.
Its {\it reduced matrices\,} or {\em marginal states\,} may be
defined by equations
\begin{eqnarray*}
\rho_A&=\sum_\alpha
a_\alpha\Tr(\rho_B^\alpha)\rho_A^\alpha:=\Tr_B(\rho_{AB}),\\
\rho_B&=\sum_\alpha
a_\alpha\Tr(\rho_A^\alpha)\rho_B^\alpha:=\Tr_A(\rho_{AB}).
\end{eqnarray*}
The reduced states $\rho_A,\rho_B$ are independent of the
decomposition (\ref{sep_decomp}) and  can be characterized
intrinsically by the following property
\begin{equation}
\langle X_A\rangle_{\rho_{AB}}=\Tr(\rho_{AB}X_A)=\Tr(\rho_A
X_A)=\langle X_A\rangle_{\rho_A},\quad\forall\quad X_A,
\end{equation}
which tells that $\rho_A$ is a ``visible" state of subsystem $A$.
This justifies the termino\-logy.



\begin{Ex}\label{Schmidt_ex} Let's identify pure state of two component system
\begin{eqnarray*}\psi=\sum_{ij}\psi_{ij}\;\alpha_i\otimes\beta_j\in\mathcal{H}_A\otimes\mathcal{H}_B
\end{eqnarray*}
with its matrix $[\psi_{ij}]$ in orthonormal bases
$\alpha_i,\beta_j$ of $\mathcal{H}_A,\mathcal{H}_B$. Then the
reduced states of $\psi$ in respective bases are given by matrices
\begin{equation}\rho_A=\psi^\dag\psi,\quad\rho_B=\psi\psi^\dag,\label{2-comp}\end{equation}
which have the same non negative spectra
\begin{equation}\label{isospect}
\mathrm{Spec}\rho_A=\mathrm{Spec}\rho_B=\lambda
\end{equation}
except extra zeros if $\dim\mathcal{H}_A\ne\dim\mathcal{H}_B$. The
isospectrality implies so called {\it Schmidt decomposition\,}
\begin{equation}\label{Schmidt}
\psi=\sum_i\sqrt{\lambda_i}\;\psi^A_i\otimes\psi^B_i,
\end{equation} where
$\psi^A_i,\psi^B_i$ are eigenvectors of $\rho_A,\rho_B$ with the
same eigenvalue $\lambda_i$.

In striking
contrast to classical case marginals of a pure state
$\psi\neq\psi_A\otimes\psi_B$ are mixed ones,
i.e. as Sr\"odinger 1935 puts it {\it``maximal knowledge of the
whole does not necessarily  includes the maximal knowledge of its
parts."}
He coined the term {\it entanglement\,} just
to describe this phenomenon.
\end{Ex}

\subsection{Statement of the  problem}
Quantum analogue  of the classical  marginal distribution
is {\it reduced state\,} $\rho_A$ of composite system
$\mathcal{H}_{AB}=\mathcal{H}_A\otimes\mathcal{H}_B$.
Accordingly, most general {\it quantum marginal problem\,} asks
about
existence of mixed state $\rho_I$ of composite system
\begin{eqnarray*}\mathcal{H}_I=\bigotimes_{i\in I}\mathcal{H}_i
\end{eqnarray*}
with given reduced states $\rho_J$ for some $J\subset I$ (cf. with
classical settings  \sref{Bell}).
Additional constraints on state $\rho_I$ may be  relevant. Here we
consider only two variations:
 \begin{itemize}
\item {\it Pure quantum marginal problem\,}

\end{itemize}
 dealing with marginals of pure state $\rho_I=|\psi\rangle\langle\psi|$, and more general
 \begin{itemize}
\item{\it Mixed quantum marginal problem\,}
\end{itemize} corresponding to a state with
given spectrum $\lambda_I=\mathrm{Spec}\,\rho_I$.

Both versions are nontrivial even for univariant margins (cf.
Example \ref{Gale}). In this case reduced states $\rho_i$ can be
diagonalized by local unitary transformations
and their compatibility depends only on spectra
$\lambda_i=\mathrm{Spec}\,\rho_i$. Note that mixed QMP say for two
component system
$\mathcal{H}_{AB}=\mathcal{H}_A\otimes\mathcal{H}_B$
is formally equivalent to the pure one for three component system
$\mathcal{H}_{AB}\otimes
\mathcal{H}_A\otimes\mathcal{H}_B$.

Pure quantum marginal problem has no classical analogue, since
projection of a point
is a point. For two component system
$\mathcal{H}_A\otimes\mathcal{H}_B$ marginal constraints amounts
to isospectrality $\mathrm{Spec}\,\rho_A=\mathrm{Spec}\,\rho_B$,
see
\Eref{isospect}. For three component system  the problem can be
stated in plain language as follows.
\begin{prob}\label{spectral} 
Let $A=[A_{ijk}]$ be complex cubic matrix and $A_1,A_2,A_3$ be
Gram matrices formed by Hermitian dot products of parallel slices
of $A$.
The question is what
are relations between spectra of matrices $A_1, A_2, A_3$?
\end{prob}
Unfortunately methods of this paper can't be applied directly to
overlapping marginals like $\rho_{AB}$, $\rho_{BC}$, $\rho_{CA}$.
\subsection{Some known results}
Here are some recent results that
laid the ground of the quantum marginal problem. They all emerge
from quantum information settings in two or three years.

\begin{thm*}[Higuchi {\etal} 2003]\label{poly} For array of qubits
$\bigotimes_{i=1}^n\mathcal{H}_i$, $\dim\mathcal{H}_i=2$ all
constraints  on margins $\rho_i$ of a pure state are given by {\it
polygonal inequalities}
\begin{eqnarray*}\lambda_i\le \sum_{j(\neq i)}\lambda_j
\end{eqnarray*}
for minimal eigenvalues $\lambda_i$ of $\rho_i$.
\end{thm*}
This result was proved independently by Sergey Bravyi who also
managed to crack mixed two qubit problem.
\begin{thm*}[Bravyi 2004]\label{2qbt}
For two cubits $\mathcal{H}_A\otimes\mathcal{H}_B$ solution of the
{\it mixed QMP} is given by  inequalities
\begin{eqnarray*}
\min(\lambda_A,\lambda_B)&\ge& \lambda_3^{AB}+\lambda_4^{AB},\\
\lambda_A+\lambda_B&\ge&
\lambda_2^{AB}+\lambda_3^{AB}+2\lambda_4^{AB}\\
|\lambda_A-\lambda_B|&\le&
\min(\lambda_1^{AB}-\lambda_3^{AB},\lambda_2^{AB}-\lambda_4^{AB}),
\end{eqnarray*}
where $\lambda_A,\lambda_B$ are minimal eigenvalues of
$\rho_A,\rho_B$ and
$\lambda^{AB}_1\ge\lambda^{AB}_2\ge\lambda^{AB}_3\ge\lambda^{AB}_4$
is spectrum of $\rho_{AB}$.
\end{thm*}

Finally for three qutrits the problem was solved by Matthias Franz
using rather advanced mathematical technology and help of a
computer. An elementary solution was found independently by
Astashi Higuchi.
\begin{thm*}[Franz 2002, Higuchi 2003]\label{franz}
All constraints on margins of a pure state of three qutrit system
$\mathcal{H}_A\otimes\mathcal{H}_B\otimes{H}_C$ are given by the
following  inequalities
\begin{eqnarray*}
\lambda_2^a+\lambda_1^a&\le&\lambda_2^b+\lambda_1^b+
\lambda_2^c+\lambda_1^c,\\
\lambda_3^a+\lambda_1^a&\le&\lambda_2^b+\lambda_1^b+
\lambda_3^c+\lambda_1^c,\\
\lambda_3^a+\lambda_2^a&\le&\lambda_2^b+\lambda_1^b+
\lambda_3^c+\lambda_2^c,\\
2\lambda_2^a+\lambda_1^a&\le&2\lambda_2^b+\lambda_1^b+
2\lambda_2^c+\lambda_1^c,\\
2\lambda_1^a+\lambda_2^a&\le&2\lambda_2^b+\lambda_1^b+
2\lambda_1^c+\lambda_2^c,\\
2\lambda_2^a+\lambda_3^a&\le&2\lambda_2^b+\lambda_1^b+
2\lambda_2^c+\lambda_3^c,\\
2\lambda_2^a+\lambda_3^a&\le&2\lambda_1^b+\lambda_2^b+
2\lambda_3^c+\lambda_2^c,
\end{eqnarray*}
where $a,b,c$ is a permutation of $A,B,C$, and the marginal
spectra are arranged in increasing order
$\lambda_1\le\lambda_2\le\lambda_3$.
\end{thm*}

Note that in contrast to the classical marginal problem, linearity
of the quantum marginal constraints is a surprising nontrivial
fact.

\subsection{Main theorem}
A general solution of the quantum marginal problem, based on
geometric invariant theory, has been found recently (Klyachko
2004).

\begin{thm}\label{mixed}
For two component system
$\mathcal{H}_{AB}=\mathcal{H}_A\otimes\mathcal{H}_B$ of format
$m\times n$ all constraints on spectra $\lambda^{AB}={\rm
Spec}\,\rho_{AB}$, $\lambda^A={\rm Spec}\,\rho_A$, $\lambda^B
={\rm Spec}\,\rho_B$
 arranged in decreasing order
are given by linear inequalities
\begin{equation}
\sum_i a_i\lambda^A_{u(i)}+\sum_j b_j\lambda^B_{\nu(j)}\le
\sum_k (a+b)_k\lambda^{AB}_{w(k)},\label{m_ineq}
\end{equation}
where $a:a_1\ge a_2\ge\cdots \ge a_m$, $b:b_1\ge b_2\ge\cdots\ge
b_n$,  $\sum a_i=\sum b_j=0$ are {\it ``test spectra"},
$a+b=\{a_i+b_j\}\!\!\downarrow$, and
$u\in S_m$, $v\in S_n$, $w\in S_{mn}$ are permutations, subject to
crucial condition $c_{uv}^w(a,b)\ne 0$ of topological nature  to
be explained below.
\end{thm}

\begin{rem}\label{tables}
Here $(a+b)_k$ denotes $k$-th term of the sequence $a_i+b_j$
arranged in decreasing order. The coefficient $c_{uv}^w(a,b)$
depends only on the order in which quantities $a_i+b_j$ appear in
the spectrum $(a+b)\!\!\downarrow$. The order changes when the
pair $(a,b)$ crosses  hyperplane
$$H_{ij|kl}:a_i+b_j=a_k+b_\ell.$$ The hyperplanes cut
the set of all pairs
$(a,b)$ into finite number of pieces called {\it cubicles}. For
each cubicle one have to check inequality (\ref{m_ineq}) only for
its {\it extremal edges}. Hence the marginal constraints amounts
to a {\it finite system} of inequalities, but
the total number of extremal edges increases rapidly. Here are
some sample data for arrays of qubits.
\begin{center}
\vspace{1mm}
\begin{tabular}{|c|c|c|c|c|c|}
\hline
\# qubits &2&3&4&5&6\\
\hline
\# edges&2&4&12&125&11344\\
\hline
\end{tabular}
\end{center}
Unfortunately for most systems the  marginal constraints are too
numerous
to be reproduced here. Therefor we give only summary table, which
shows how complicate may be the answer.
\vspace{3mm}
\begin{center}
\begin{tabular}{|c|c|c|}
\hline\label{table}
System & Rank&\#Inequalities\\
\hline
${2\times2}$&${2}$&${7}$\\
\hline
${2\times2\times2}$&${3}$&${40}$\\
${2\times3}$&${3}$&${41}$\\
\hline
${2\times4}$&${4}$&${234}$\\
${3\times3}$&${4}$&${387}$\\
${2\times2\times3}$&${4}$&${442}$\\
${2\times2\times2\times2}$&${4}$&${805}$\\
\hline
\end{tabular}
\end{center}
\end{rem}

\subsection{Hidden geometry and topology}\label{topology}
Here we explain the meaning of the coefficient $c_{uv}^w(a,b)$ in
statement of the theorem and show how they can be calculated.
Consider set of Hermitian operators
$X_A:\mathcal{H}_A\rightarrow\mathcal{H}_A$ with given spectrum
$\mathrm{Spec}(X_A)=a$ and call it {\it flag variety\,}
\begin{eqnarray*}\mathcal{F}_a(\mathcal{H}_A):=\{X_A\mid\mathrm{Spec}(X_A)=a\}.
\end{eqnarray*}
For two flag varieties $\mathcal{F}_a(\mathcal{H}_A)$ and
$\mathcal{F}_b(\mathcal{H}_B)$ define map
\begin{eqnarray*}\varphi_{ab}:
\mathcal{F}_a(\mathcal{H}_A)\times\mathcal{F}_b(\mathcal{H}_B) &\longrightarrow&
\mathcal{F}_{a+b}(\mathcal{ H}_A\otimes \mathcal{ H}_B),\\
\qquad\qquad X_A\times X_B&\longmapsto& \;\;\, X_A\otimes 1+1\otimes X_B.
\end{eqnarray*}
The coefficients $c^w_{uv}(a,b)$ come from the  induced morphism
of cohomology
\begin{eqnarray*}\varphi_{ab}^*: H^*(\mathcal{F}_{a+b}(\mathcal{ H}_{AB}))\rightarrow
H^*(\mathcal{F}_{a}(\mathcal{H}_{A}))\otimes
H^*(\mathcal{F}_{b}(\mathcal{H}_B))
\end{eqnarray*} written in the basis of {\it
Schubert cocycles} $\sigma^w$ 
\begin{eqnarray*}\varphi^*_{ab}:\sigma^w\mapsto
\sum_{u,v} c^w_{uv}(a,b)\sigma^u\otimes\sigma^v.
\end{eqnarray*}
We'll give  below an algorithm for their calculation. For this we
need another description of the cohomology of flag varieties due
to (Bernstein {\etal} 1973). Specifically, for {\it simple\,}
spectrum $a$ eigenspaces of operator
$X_A\in\mathcal{F}_a(\mathcal{H}_A)$ of given eigenvalue $a_i$
form a line bundle $\mathcal{L}_i^A$ on
$\mathcal{F}_a(\mathcal{H}_A)$. Their Chern classes
$x_i^A=c_1(\mathcal{L}_i^A)$ generate cohomology ring
$H^*(\mathcal{F}_a(\mathcal{H}_A))$ and in this setting morphism
$\varphi_{ab}^*$ admits a simple description
\begin{eqnarray*}\varphi^*_{ab}(\mathcal{L}_k^{AB})=\mathcal{L}^A_i\boxtimes\mathcal{L}^B_j,\quad
\varphi^*_{ab}:x_k^{AB}\mapsto x_i^{A}+x_j^{B}
\end{eqnarray*}
where $(a+b)_k=a_i+b_j$. In terms of the {\it canonical
generators\,} $x_i=c_1(\mathcal{L}_i)$ Schubert cocycle $\sigma^w$
is given by {\it Schubert polynomial} (Macdonald 1991)
\begin{eqnarray*}S_w(x_1,x_2,\ldots)=\partial_{w^{-1}w_0}(x_1^{n-1}x_2^{n-2}\cdots
x_{n-1}),
\end{eqnarray*}
where  $w\in S_n$ is a permutation of $1,2,\ldots,n$,
$w_0=(n,n-1,\ldots,2,1)$ reversion of the order, operator
$\partial_w=\partial_{i_1}\partial_{i_2}\cdots\partial_{i_\ell}$,
is defined via minimal decomposition $w=s_{i_1}s_{i_2}\cdots
s_{i_\ell}$  into product of transpositions $s_i=(i,i+1)$,
$\ell=\ell(w)$ is the number of inversion of $w$ called its {\it
length\,}, and finally
\begin{eqnarray*}
\partial_if=
\frac{f(\ldots,x_i,x_{i+1},\ldots)-f(\ldots,x_{i+1},x_i,\ldots)}
{x_i-x_{i+1}}
\end{eqnarray*}
is finite difference operator.
This leads to the {\it computational formula}
\begin{equation}\label{coef2comp}c_{uv}^w(a,b)=\partial^A_u\partial^B_v
S_w(x^{AB})\left|_{x^{AB}_k=x^A_i+x^B_j}\right.
\end{equation} where
$(a+b)_k=a_i+b_j$ and $\ell(w)=\ell(u)+\ell(v)$. It can be easily
implemented into a computer progrm. Recall that in order to get a
finite system of inequalities one have also to find all the
extremal edges and use them as the test spectra $(a,b)$.
\begin{Ex}
Note that for identical permutations $u,v,w$ the coefficient
$c_{uv}^w(a,b)$ is equal to 1.
Hence inequality 
\begin{eqnarray*}
\sum_i a_i\lambda^A_{i}+\sum_j b_j\lambda^B_{j}\le \sum_{k}
(a+b)_k\lambda^{AB}_{k}
\end{eqnarray*}
holds for all test spectra $(a,b)$.
This amounts to {finite} system of {\it basic inequalities\,} (Han
\etal 2004)
\begin{eqnarray*}\lambda^A_1+\lambda^A_2+\cdots+\lambda^A_k&\le&
\lambda^{AB}_1+\lambda^{AB}_2+\cdots+\lambda^{AB}_{kn},\quad k\le m,
\\
\lambda^B_1+\lambda^B_2+\cdots+\lambda^B_\ell&\le&
\lambda^{AB}_1+\lambda^{AB}_2+\cdots+\lambda^{AB}_{m\ell },\quad\ell\le n.
\end{eqnarray*}
\end{Ex}

\subsection{Array of qubits}
The calculations can be essentially reduced using the following
result.

\begin{thm} In the setting of Theorem \ref{mixed} all marginal constraints are
given by inequalities (\ref{m_ineq}) with $c_{uv}^w(a,b)=1$.
\end{thm}

\begin{Ex}
One can check that for array of qubits
$\bigotimes_{i=1}^n\mathcal{H}_i$ all inequalities \eref{m_ineq}
with {\it odd\,} coefficient $c^{w}_{u_1u_2\ldots
u_n}(a_1,a_2,\ldots,a_n)$ can be obtained from the basic
inequality
$$\sum_i a_i(\lambda^{(i)}_1-\lambda^{(i)}_2)\le \sum_{\pm}(\pm a_1\pm a_2\pm\cdots\pm
a_n)_k\lambda_k$$ by transposition
$\lambda_k\leftrightarrow\lambda_{k+1}$, $k=\mbox{odd}$, combined
with sign change of a  summand in LHS. Here $a_i\ge -a_i$ are the
{\it test spectra}, $\lambda$ and $\lambda^{(i)}$ are spectra of
state $\rho$ of the system and its one qubit reduced state
$\rho^{(i)}$ respectively.  To get a finite system one has only to
find the extremal edges. For large  $N$ this may be a challenge,
see Remark \ref{tables}.
\end{Ex}
\begin{Ex}\label{3qbt} For 3-qubit the theorem
returns the following list of marginal inequalities grouped by
their extremal edges. The first inequality in each group is the
basic one. The transposed eigenvalues in modified inequalities
typeset in bold face. We expect
$\Delta_i=\lambda^{(i)}_1-\lambda^{(i)}_2$ to be arranged in
increasing order $\Delta_1\le\Delta_2\le\Delta_3$.
$$
\begin{array}{rcl}
\Delta_3
&\le&\lambda_1+\lambda_2+\lambda_3+\lambda_4-\lambda_5-\lambda_6-\lambda_7-\lambda_8.\\[2mm]
\Delta_2+\Delta_3 &\le& 2\lambda_1+ 2\lambda_2- 2\lambda_7- 2\lambda_8.\\[2mm]
\Delta_1+\Delta_2+\Delta_3&\le&3\lambda_1+\lambda_2+\lambda_3+\lambda_4-\lambda_5-\lambda_6-\lambda_7-3\lambda_8,\\
-\Delta_1+\Delta_2+\Delta_3&\le&3{\boldsymbol{\lambda_2}}+{\boldsymbol{\lambda_1}}+\lambda_3+\lambda_4-\lambda_5-\lambda_6-\lambda_7-3\lambda_8,\\
-\Delta_1+\Delta_2+\Delta_3&\le&3\lambda_1+\lambda_2+\lambda_3+\lambda_4-\lambda_5-\lambda_6-
{\boldsymbol{\lambda_8}}-3{\boldsymbol{\lambda_7}}.
\\[2mm]
\Delta_1+\Delta_2+2\Delta_3&\le&4\lambda_1+2\lambda_2+2\lambda_3-2\lambda_6-2\lambda_7-4\lambda_8,\\
-\Delta_1+\Delta_2+2\Delta_3&\le&4{\boldsymbol{\lambda_2}}+2{\boldsymbol{\lambda_1}}+2\lambda_3-2\lambda_6-2\lambda_7-4\lambda_8,\\
-\Delta_1+\Delta_2+2\Delta_3&\le&4\lambda_1+2\lambda_2+2{\boldsymbol{\lambda_4}}-2\lambda_6-2\lambda_7-4\lambda_8,\\
-\Delta_1+\Delta_2+2\Delta_3&\le&4\lambda_1+2\lambda_2+2\lambda_3-2{\boldsymbol{\lambda_5}}-2\lambda_7-4\lambda_8,\\
-\Delta_1+\Delta_2+2\Delta_3&\le&4\lambda_1+2\lambda_2+2\lambda_3-2\lambda_6-2{\boldsymbol{\lambda_8}}-
4{\boldsymbol{\lambda_7}}.
\end{array}
$$
\end{Ex}
\section{$\mathbf{N}$-representability problem}
\subsection{Physical background}

The quantum marginal problem may be complicated by additional
constraints on state $\psi$. For example, Pauli principle implies
that state space of $N$ identical particles shrinks to symmetric
tensors $S^N\mathcal{H}\subset\mathcal{H}^{\otimes N}$ for bosons
and to skew symmetric tensors $\wedge^N\mathcal{H}$ for fermions.
For such systems reduced density matrices appear in the second
quantization formalism in the form
$$\begin{array}{ccccl}\rho^{(1)}&=&\langle\psi|a_i^\dag
a_j|\psi\rangle&=&\mbox{1 particle
RDM,}\\
\rho^{(2)}&=&\langle\psi|a_i^\dag a_j^\dag a_ka_l|\psi\rangle &=&\mbox{2 particle
RDM, etc.}
\end{array}$$
Their physical importance stems from the observation that, say for
{\it fermionic system}, like multi electron atom or molecule, with
pairwise interaction
\begin{eqnarray*}H=\sum_i^N H_i+\sum_{i<j}H_{ij}
\end{eqnarray*} the energy
of state $\psi$ depends only on 2-point RDM
\begin{eqnarray*}E={N\choose 2}{\rm Tr}\,(H^{(2)}\rho^{(2)}),
\end{eqnarray*}
where $H^{(2)}=\frac{1}{N-1}[H_1+H_2]+H_{12}$ is reduced two
particle Hamiltonian. This allows, for example, to express the
energy of ground state $E_0$ via 2-point RDM
\begin{eqnarray*}E_0={N\choose 2}\min_{\rho^{(2)}=\mbox{RDM}}{\rm Tr}(H^{(2)}\rho^{(2)}).
\end{eqnarray*}
The problem however  is that
it is not obvious what conditions the RDM itself
should satisfy. This is what the quantum marginal problem  is
about. In this settings it
is known   from early sixties as {\it $N$- representability
problem\,} (Coulson 1960, Coleman 1963). Later in mid 90-th the
problem was regarded as one of ten most prominent research
challenges in quantum chemistry (Stillinger \etal 1995).
Its solution allows to calculate nearly all properties of matter
which are of interest to chemists and physicists. For current
state of affairs and more history see (Coleman and Yukalov 2000,
Coleman 2001).

\subsection{One point $N$-representability}\label{one_pnt}
Here we
outline a solution of the problem for one point reduced states.
Following chemists we treat them
as electron  density and accordingly use normalization ${\rm
Tr}\;\rho^{(1)}=N$ while keeping $\Tr\rho=1$.
There are few cases where complete
solution of one point $N$-representability  was known prior 2005:
\begin{itemize}
\item Pauli principle: $0\le \lambda_i\le 1$, \quad
$\lambda={\rm Spec}\;\rho^{(1)}$. This condition provides a
criterion  for {\it mixed $N$-representability} (Coleman 1963).
\item Criterion for pure $N$-representability for two particles $\wedge^2\mathcal{H}_r$ or two holes
$\wedge^{r-2}\mathcal{H}_r$ is given by even degeneration of all
eigenvalues of $\rho^{(1)}$, except $0$ (resp. $1$) for odd
$r=\dim\mathcal{H}_r$ (Coleman 1963).
\item 
For system of three fermions of dimension six
$\wedge^3\mathcal{H}_6$ all constraints on one point reduced
matrix of a pure state are given by the following (in)equalities
\begin{eqnarray}\label{B-D1}
\lambda_1+\lambda_6=\lambda_2+\lambda_5=\lambda_3+\lambda_4=1,\quad
\lambda_4\le \lambda_5+\lambda_6,
\end{eqnarray} where
$\lambda_1\ge\lambda_2\ge\lambda_3\ge\lambda_4\ge\lambda_5\ge\lambda_6$
is spectrum of $\rho^{(1)}$.
\end{itemize}
The last result belongs to Borland and Dennis 1972 who commented
it as follows:
\begin{quote}
{\it ``We have no apology for consideration of such a special
case. The general $N$-representability problem is so difficult and
yet so fundamental for many branches  of science that each
concrete result is useful in shedding light on the nature of
general solution."}
\end{quote}
For more then 30 years passed after this theorem no other solution
of $N$-representability problem has been found. Borland and Dennis
derived  their criterion from an extensive computer experiment,
and later proved it with help provided by M.B.~Ruskai and
R.L.~Kingsly. They also conjectured solutions  for systems
$\wedge^3\mathcal{H}_7,
\wedge^4\mathcal{H}_7,\wedge^4\mathcal{H}_8$, e.g. for
$\wedge^3\mathcal{H}_7$ one point pure representability  is given
by 4 inequalities
\begin{eqnarray}\label{B-D}
\begin{array}{cc}
\lambda_1+\lambda_6+\lambda_7\ge 1,&\lambda_2+\lambda_5+\lambda_7\ge
1,\\
\lambda_3+\lambda_4+\lambda_7\ge 1,&\lambda_3+\lambda_5+\lambda_6\ge
1,
\end{array}
\end{eqnarray}
but they failed to prove them. The  conjectures turn out to be
true and covered by the following general result. 

\begin{thm}\label{mix_frm_thm}
For mixed state $\rho$ of\, $n$\,-\,fermion system
$\wedge^n\mathcal{H}_r$ all constraints on spectra $\nu={\rm
Spec}\,\rho$ and $\lambda={\rm Spec}\,\rho^{(1)}$ are given by
inequalities
\begin{equation}\label{mix_frm_inq} \sum_i a_i\lambda_{v(i)}\le
\sum_j (\wedge^n a)_j\nu_{w(j)}
\end{equation} for all ``test spectra" $a: a_1\ge a_2\ge\cdots
\ge a_r$, $\sum a_i =0$. Here $\wedge^n
a=\{a_{i_1}+a_{i_2}+\cdots+a_{i_n}\}\downarrow$ consists of all
sums $a_{i_1}+a_{i_2}+\cdots+a_{i_n}$, $i_1<i_2<\cdots<i_n$
arranged in decreasing order, and $v\in S_r,w\in S_{n\choose r}$
are permutations subject to topological condition $c_w^v(a)\neq 0$
to be explained below.
\end{thm}
\begin{rem}
Recall that the spectra $\lambda$ and $\nu$ are arranged in
decreasing order and normalized to trace  $n$ and $1$
respectively.
 Similarly to Theorem \ref{mixed} the coefficients
$c_w^v(a)$ are defined via
{\it flag variety}
$\mathcal{F}_a(\mathcal{H}):=\{A:\mathcal{H}\rightarrow\mathcal{H}\mid{\rm
Spec}(A)=a\}$ and morphism
\begin{eqnarray*}\varphi_a:
\mathcal{F}_a(\mathcal{H})&\rightarrow&
\mathcal{F}_{\wedge^na}(\wedge^n\mathcal{ H})\\
\qquad\quad A\quad&\mapsto& \quad A^{(n)}
\end{eqnarray*}
where operator $A^{(n)}:\wedge^n\mathcal{H}\rightarrow
\wedge^n\mathcal{H}$ acts as differential
$$A^{(n)}:x_1\wedge x_2\wedge\ldots\wedge x_n\mapsto
\sum_i x_1\wedge x_2\wedge\ldots\wedge Ax_i\wedge\ldots\wedge
x_n.$$
The coefficients $c_w^v(a)$ come from the induced morphism of
cohomology
\begin{eqnarray*}\varphi_a^*: H^*(\mathcal{F}_{\wedge^na}(\wedge^n\mathcal{ H}))\rightarrow
H^*(\mathcal{F}_{a}(\mathcal{H}))
\end{eqnarray*} written in the basis of {\it
Schubert cocycles} $\sigma_w$
\begin{eqnarray*}\varphi^*_a:\sigma_w\mapsto
\sum_v c_w^v(a)\sigma_v.
\end{eqnarray*}
They can be calculated
by equation
\begin{eqnarray*}c_w^v(a)=\partial_vS_w(z)\left|_{z_k=x_{i_1}+x_{i_2}+\cdots+x_{i_n}}\right.,
\end{eqnarray*}
where  $a_{i_1}+a_{i_2}+\cdots+a_{i_n}$ is $k$-th term of the
sequence $\wedge^n a$, cf.
\sref{topology}.
\end{rem}

\begin{Ex}\label{Riem_Ric1} For system
$\wedge^2\mathcal{H}_4$ the marginal constraints on
$\nu=\mathrm{Spec }\,\rho$ and $\lambda=\mathrm{Spec}\,\rho^{(1)}$
are given by inequalities
\begin{eqnarray}\label{Rim_Ric_inq}
\hspace{30pt}2\lambda_1&\le \nu_1+\nu_2+\nu_3\nonumber\\
\hspace{30pt}2\lambda_4&\ge \nu_4+\nu_5+\nu_6\nonumber\\
\hspace{-5pt}2(\lambda_1-\lambda_4)&\le
\nu_1+\nu_2-\nu_5-\nu_6\nonumber\\
\hspace{-44pt}\lambda_1+\lambda_2-\lambda_3-\lambda_4&\le \nu_1-\nu_6\\
\hspace{-44pt}\lambda_1-\lambda_2+\lambda_3-\lambda_4&\le\min(\nu_1-\nu_5,\nu_2-\nu_6)\nonumber\\
\hspace{-50pt}|\lambda_1-\lambda_2-\lambda_3+\lambda_4|&\le\min(\nu_1-\nu_4,\nu_2-\nu_5,
\nu_3-\nu_6)\nonumber\\
\fl2\max(\lambda_1-\lambda_3,\lambda_2-\lambda_4)&\le\min(\nu_1+\nu_3-\nu_5-\nu_6,
\nu_1+\nu_2-\nu_4-\nu_6)\nonumber\\
\fl 2\max(\lambda_1-\lambda_2,\lambda_3-\lambda_4)&\le\min(\nu_1+\nu_3-\nu_4-\nu_6,
\nu_2+\nu_3-\nu_5-\nu_6,\nonumber
\nu_1+\nu_2-\nu_4-\nu_5).
\end{eqnarray}
In these inequalities  we use standard normalization
$\Tr\rho=\Tr\rho^{(1)}$. Recall that all spectra are written in
decreasing order.
\end{Ex}
\begin{Ex} Similar compatibility conditions for system
$\wedge^2\mathcal{H}_5$ contain {460} independent inequalities
which can't be reproduced here.
\end{Ex}
\begin{rem} \label{Riem_Ric} In tensor algebra the reduction
$\rho_{AB}\mapsto\rho_A$ is known as {\it contraction\,}. Most
mathematicians are familiar with this procedure from differential
geometry, where say Ricci curvature tensor
$\mathrm{Ric}:\mathcal{T}\rightarrow\mathcal{T}$ is defined as
contraction of Riemann curvature
$R:\wedge^2\mathcal{T}\rightarrow\wedge^2\mathcal{T}$ (here
$\mathcal{T}$ stands for tangent bundle). This allows  to
interpret inequalities (\ref{Rim_Ric_inq}) as relations between
spectra of Riemann and Ricci curvatures of four-manifold.
Recall that in general relativity Ricci curvature
is governed by energy-momentum tensor, i.e.  by physical content
of the space, while Riemann curvature is responsible for its
geometry and topology. The above constraints impose some bounds on
influence of matter on geometry.
\end{rem}


\subsection{Pure $N$-representability in dimension $\le 8$}
Here I'll give  an account of our joint work with Murat
Altunbudak. The details will be published elsewhere.

Formally solution of pure marginal problem can be deduced from
inequalities
\eref{mix_frm_inq} of Theorem \ref{mix_frm_thm} by putting
$\nu_i=0$ for  $i\ne1$. However for a system like
$\wedge^4\mathcal{H}_8$
we confront with an immense symmetric group of degree ${8\choose
4}=70$.
A representation theoretical interpretation of
$N$-representability discussed below allows to circumvent
this difficulty.

Let's start with decomposition of a symmetric power of
$\wedge^n\mathcal{H}$,  called {\it plethysm\,},
into irreducible components
\begin{equation}\label{pleth_dec}
S^m(\wedge^n\mathcal{H})=\sum_\lambda m_\lambda\mathcal{H}_\lambda
\end{equation}
of the unitary group $\mathrm{U}(\mathcal{H})$. The components
$\mathcal{H}_\lambda$, entering into the decomposition with some
multiplicities $m_\lambda\ge0$, can be
parameterized
by {\it Young diagrams\,}
\begin{equation*}\lambda:
\lambda_1\ge\lambda_2\ge
\cdots\ge
\lambda_r\ge 0
\end{equation*}
of size $|\lambda|=\sum_i\lambda_i=n\cdot m$ that fit into
$r\times m$ rectangular, $r=\dim\mathcal{H}$. It is instructive to
treat the diagrams as {\it spectra\,}. We are interested in
asymptotic of these spectra as $m\rightarrow\infty$ and therefor
normalize them to a fixed size $\widetilde{\lambda}=\lambda/m$,
$\Tr\widetilde{\lambda}=n$.

\begin{thm}\label{pleth} Every
$\widetilde{\lambda}$ obtained from irreducible component
$\mathcal{H}_\lambda\subset S^d(\wedge^n\mathcal{H})$ is spectrum
one point reduced matrix $\rho^{(1)}$ of a pure state $\psi\in
\wedge^n\mathcal{H}$. Moreover every one point reduced spectrum is
a convex combination of the spectra $\widetilde{\lambda}$ with
bounded $m\le M$.
\end{thm}

A similar result holds in standard settings of the quantum
marginal problem (Franz 2002, Christandl and Mitchison 2004,
Klyachko 2004, Christandl \etal 2005).

Note that representation $S^m(\wedge^{r-n}\mathcal{H})$ is dual to
$S^m(\wedge{^n}\mathcal{H})$ and hence
\begin{equation}
S^m(\wedge^{r-n}\mathcal{H})=\sum_\lambda
m_\lambda\mathcal{H}_{\lambda^*},
\end{equation}
where $\lambda^*$ is complement of the diagram $\lambda$ to the
rectangle  $r\times m$ and the multiplicity $m_\lambda$ is the
same as in \eref{pleth_dec}. Thus we arrived at the following {\it
particle-hole duality\,}.
\begin{cor} Marginal constraints on spectrum of reduced matrix of
a pure state for system $\wedge^{r-n}\mathcal{H}_r$ can be
obtained from that of the system $\wedge^n\mathcal{H}_r$ by
substitution $\lambda_i\mapsto 1-\lambda_{r+1-i}$.
\end{cor}

\begin{Ex} There are few cases where decomposition
\eref{pleth_dec} is explicitly known, for example
\begin{equation*}S^m(\wedge^2\mathcal{H}_r)=\sum_{|\lambda|=2m,\;\lambda=\mathrm{even}}
\mathcal{H}_\lambda,
\end{equation*}
where the sum is extended over diagrams $\lambda\subset r\times m$
with even multiplicity of every nonzero row (Macdonald 1995).
Together with theorem \ref{pleth} and  the particle-hole duality
this implies Coleman's criteria of pure N-representability for
systems of two particles $\wedge^2\mathcal{H}_r$ and two holes
$\wedge^{r-2}\mathcal{H}_r$ mentioned at the beginning of section
\ref{one_pnt}.
\end{Ex}
\begin{Ex}
Borland-Dennis equations (\ref{B-D1}) imply that every component
$\mathcal{H}_\lambda\subset S^m(\wedge^3\mathcal{H}_6)$ is
selfdual $\lambda=\lambda^*$. It seems mathematicians missed this
fact, which holds only for this specific system. Note that wedge
product ensure selfduality of  $\wedge^3\mathcal{H}_6$ and hence
of the plethysm $S^m(\wedge^3\mathcal{H}_6)$. However apparently
there is no simple way to extend this to every component
$\mathcal{H}_\lambda \subset S^m(\wedge^3\mathcal{H}_6)$.
\end{Ex}

Theorem \ref{pleth} for any fixed $M$ gives an inner approximation
to the set of all possible reduced spectra, while any set of
inequalities (\ref{mix_frm_inq}) of theorem \ref{mix_frm_thm}
amounts to its outer approximation. This suggests the following
approach to pure $N$-representability problem, which combines both
theorems.
\begin{itemize}
\item Find all irreducible components $\mathcal{H}_\lambda\subset S^m(\wedge^n\mathcal{H})$ for $m\le M$ starting with $M=1$.
\item Calculate convex hull of the corresponding reduced spectra
$\widetilde\lambda$.
\item Check whether or not all inequality defining facets of the convex
hull fit into the form  \eref{mix_frm_inq} of Theorem
\ref{mix_frm_thm}.
\item If they do then all inequalities are found.
Otherwise increase $M\mapsto M+1$.
\end{itemize}
\begin{rem}
The success of this approach depends on the degrees  of generators
of the module of covariants of the system $\wedge^n\mathcal{H}_r$.
Generically the degrees are expected to be huge as well as the
whole number of the resulting inequalities. However for systems of
rank $r\le 8$ and for $r=9,\; n\ne 4,5$ the module of covariants
is free (Vinberg and Popov 1992) and the degrees of the generators
should be reasonably small.
\end{rem}

Indeed an inexpensive PC, assisted with some dirty tricks, managed
to resolve $N$-representability problem for rank $r\le 8$.
Recall that for two fermions or two holes the answer is known, see
\sref{one_pnt}. Together with  particle-hole duality this bound us
to the range $3\le n\le r/2$. The corresponding
constraints are
listed below. They are grouped by the extremal edges and
use chemical normalization $\sum_i\lambda_i=n$.

\begin{itemize}
\item
$\wedge^3\mathcal{H}_6$.
\begin{equation*}
\lambda_1+\lambda_6=\lambda_2+\lambda_5=\lambda_3+\lambda_4=1,\quad
\lambda_4\le \lambda_5+\lambda_6
\end{equation*}
\item
$\wedge^3\mathcal{H}_7$.
\begin{eqnarray*}\label{7-3-2}
\hspace{-10pt}-4\lambda_1+3\lambda_2+3\lambda_3+3\lambda_4+3\lambda_5-4\lambda_6-4\lambda_7\leq 2\\
3\lambda_1-4\lambda_2+3\lambda_3+3\lambda_4-4\lambda_5+3\lambda_6-4\lambda_7\leq 2\\
3\lambda_1+3\lambda_2-4\lambda_3-4\lambda_4-3\lambda_5+3\lambda_6-4\lambda_7\leq 2\\
3\lambda_1+3\lambda_2-4\lambda_3+3\lambda_4-4\lambda_5-4\lambda_6+3\lambda_7\leq
2\nonumber
\end{eqnarray*}
\item
$\wedge^3\mathcal{H}_8$.
\begin{eqnarray*} 3 \lambda_1 - \lambda_2 - \lambda_3 - \lambda_4 -
\lambda_5 - \lambda_6 - \lambda_7 + 3 \lambda_8 \leq 1
\end{eqnarray*}
\begin{eqnarray*}
\hspace{-10pt}-\lambda_1 + \lambda_2 + \lambda_3 + \lambda_4 + \lambda_5 -\lambda_6 - \lambda_7 - \lambda_8 &\leq1\\
\lambda_1 + \lambda_2 - \lambda_3 - \lambda_4 + \lambda_5 +\lambda_6 - \lambda_7 - \lambda_8 &\leq 1\\
\lambda_1 + \lambda_2 - \lambda_3 + \lambda_4 - \lambda_5 -\lambda_6 + \lambda_7 - \lambda_8 &\leq 1\\
\lambda_1 - \lambda_2 + \lambda_3 + \lambda_4 - \lambda_5
+\lambda_6 - \lambda_7 - \lambda_8 &\leq 1
\end{eqnarray*}
\begin{eqnarray*}
2 \lambda_1 + \lambda_2 - 2 \lambda_3 - \lambda_4 - \lambda_6+ \lambda_8 &\leq1\\
2 \lambda_1 - \lambda_2 - \lambda_4 + \lambda_6 - 2 \lambda_7+ \lambda_8 &\leq 1 \\
\lambda_3 + 2 \lambda_4 - 2 \lambda_5 -\lambda_6 -\lambda_7 +\lambda_8 &\leq 1 \\
\lambda_1 + 2 \lambda_2 - 2 \lambda_3 - \lambda_5-\lambda_6+\lambda_8 &\leq 1\\
2\lambda_1-\lambda_2+\lambda_4-2\lambda_5-\lambda_6+\lambda_8&\leq1\\
\end{eqnarray*}
\begin{eqnarray*}
5 \lambda_1 + 5 \lambda_2 - 7 \lambda_3 - 3 \lambda_4 -
3\lambda_5 + \lambda_6 + \lambda_7 + \lambda_8 &\leq 3\\
5\lambda_1 - 3 \lambda_2 - 3 \lambda_3 + \lambda_4 + \lambda_5 + 5
\lambda_6 - 7 \lambda_7 + \lambda_8 &\leq 3
\end{eqnarray*}
\begin{eqnarray*}
5 \lambda_1 + \lambda_2 - 3 \lambda_3 + \lambda_4 - 3
\lambda_5 + \lambda_6 - 3 \lambda_7 + \lambda_8 &\leq 3\\
\lambda_1 + \lambda_2 + \lambda_3 + 5 \lambda_4 - 3 \lambda_5
- 3 \lambda_6 - 3 \lambda_7 + \lambda_8 &\leq 3\\
\lambda_1 + 5 \lambda_2 - 3 \lambda_3 + \lambda_4 + \lambda_5 - 3
\lambda_6 - 3 \lambda_7 + \lambda_8 &\leq 3
\end{eqnarray*}
\begin{eqnarray*}
9 \lambda_1 + \lambda_2 - 7 \lambda_3 - 7 \lambda_4 - 7
\lambda_5 + \lambda_6 + \lambda_7 + 9 \lambda_8 &\leq 3\\
9\lambda_1 - 7 \lambda_2 - 7 \lambda_3 + \lambda_4 + \lambda_5 +
\lambda_6 - 7 \lambda_7 + 9 \lambda_8 &\leq 3\\
\end{eqnarray*}
\begin{eqnarray*}
7 \lambda_1 - \lambda_2 - \lambda_3 - \lambda_4 - \lambda_5 +
7 \lambda_6 - 9 \lambda_7 - \lambda_8 &\leq 5 \\
7 \lambda_1 - \lambda_2 - \lambda_3 + 7 \lambda_4 - 9
\lambda_5 - \lambda_6 - \lambda_7 - \lambda_8 &\leq 5 \\
7 \lambda_1 + 7 \lambda_2 - 9 \lambda_3 - \lambda_4 -
\lambda_5 - \lambda_6 - \lambda_7 - \lambda_8 &\leq 5\\
\hspace{-10pt}-\lambda_1 - \lambda_2 + 7 \lambda_3 + 7 \lambda_4 -
\lambda_5 - \lambda_6 - 9 \lambda_7 - \lambda_8 &\leq 5\\
\hspace{-10pt}-\lambda_1 + 7 \lambda_2 - \lambda_3 + 7 \lambda_4 -
\lambda_5 - 9 \lambda_6 - \lambda_7 - \lambda_8 &\leq 5\\
\hspace{-10pt}-\lambda_1 + 7 \lambda_2 - \lambda_3 - \lambda_4 + 7 \lambda_5 -
\lambda_6 - 9 \lambda_7 - \lambda_8 &\leq 5
\end{eqnarray*}
\begin{eqnarray*}
\hspace{-10pt}-3 \lambda_1 + 5 \lambda_2 + 5 \lambda_3 + 13 \lambda_4 - 11
\lambda_5 - 3 \lambda_6 - 11 \lambda_7 + 5 \lambda_8 &\leq 7 \\
5 \lambda_1 + 13 \lambda_2 - 11 \lambda_3 + 5 \lambda_4 - 11
\lambda_5 - 3 \lambda_6 - 3 \lambda_7 + 5 \lambda_8 &\leq 7 \\
5 \lambda_1 - 3 \lambda_2 + 5 \lambda_3 + 13 \lambda_4 - 11
\lambda_5 - 11 \lambda_6 - 3 \lambda_7 + 5 \lambda_8 &\leq 7 \\
5 \lambda_1 + 13 \lambda_2 - 11 \lambda_3 - 3 \lambda_4 + 5
\lambda_5 - 11 \lambda_6 - 3 \lambda_7 + 5 \lambda_8 &\leq 7
\end{eqnarray*}
\begin{eqnarray*}
19 \lambda_1 + 11 \lambda_2 - 21 \lambda_3 - 13 \lambda_4 - 5\lambda_5 - 5 \lambda_6 + 3 \lambda_7 + 11 \lambda_8 &\leq 9\\
19 \lambda_1 - 13 \lambda_2 - 5 \lambda_3 - 5 \lambda_4 + 3\lambda_5 + 11 \lambda_6 - 21 \lambda_7 + 11 \lambda_8 &\leq9\\
11 \lambda_1 + 19 \lambda_2 - 21 \lambda_3 - 5 \lambda_4 - 13\lambda_5 - 5 \lambda_6 + 3 \lambda_7 + 11 \lambda_8 &\leq 9\\
\hspace{-10pt}-5 \lambda_1 + 3 \lambda_2 + 11 \lambda_3 + 19 \lambda_4 - 21
\lambda_5 - 13 \lambda_6 - 5 \lambda_7 + 11 \lambda_8 &\leq9
\end{eqnarray*}
\item
$\wedge^4\mathcal{H}_8$.
\begin{eqnarray*}\label{edge1}
5\lambda_1+\lambda_2+\lambda_3-3\lambda_4+\lambda_5-3\lambda_6-3\lambda_7+\lambda_8&\leq4\\
\lambda_1+\lambda_2+5\lambda_3-3\lambda_4+\lambda_5+\lambda_6-3\lambda_7-3\lambda_8&\leq4\\
\lambda_1+\lambda_2+\lambda_3+\lambda_4+5\lambda_5-3\lambda_6-3\lambda_7-3\lambda_8&\leq4\\
\lambda_1+5\lambda_2+\lambda_3-3\lambda_4+\lambda_5-3\lambda_6+\lambda_7-3\lambda_8&\leq4\\
5\lambda_1-3\lambda_2+\lambda_3+\lambda_4+\lambda_5+\lambda_6-3\lambda_7-3\lambda_8&\leq4\\
5\lambda_1+\lambda_2+\lambda_3-3\lambda_4-3\lambda_5+\lambda_6+\lambda_7-3\lambda_8&\leq4\\
5\lambda_1+\lambda_2-3\lambda_3+\lambda_4+\lambda_5-3\lambda_6+\lambda_7-3\lambda_8&\leq4
\end{eqnarray*}
\begin{eqnarray*}
\hspace{-10pt}-\lambda_1+3\lambda_2+3\lambda_3-\lambda_4+3\lambda_5-\lambda_6-\lambda_7-5\lambda_8&\leq4\\
3\lambda_1+3\lambda_2-\lambda_3-\lambda_4+3\lambda_5-5\lambda_6-\lambda_7-\lambda_8&\leq4\\
3\lambda_1+3\lambda_2+3\lambda_3-5\lambda_4-\lambda_5-\lambda_6-\lambda_7-\lambda_8&\leq4\\
3\lambda_1-\lambda_2+3\lambda_3-\lambda_4+3\lambda_5-\lambda_6-5\lambda_7-\lambda_8&\leq4\\
3\lambda_1+3\lambda_2-\lambda_3-\lambda_4-\lambda_5-\lambda_6+3\lambda_7-5\lambda_8&\leq4\\
3\lambda_1-\lambda_2-\lambda_3+3\lambda_4+3\lambda_5-\lambda_6-\lambda_7-5\lambda_8&\leq4\\
3\lambda_1-\lambda_2+3\lambda_3-\lambda_4-\lambda_5+3\lambda_6-\lambda_7-5\lambda_8&\leq4
\end{eqnarray*}
\end{itemize}
\begin{rem}
The marginal inequalities are independent and written in the form
(\ref{mix_frm_inq}) of theorem \ref{mix_frm_thm}. Using the
normalization equation $\Tr\rho=n$ they can be transformed  in
many different ways. For example, the above constraints for system
$\wedge^3\mathcal{H}_7$ are equivalent to  inequalities
\eref{B-D}. The inequalities for $\wedge^4\mathcal{H}_8$ can be
recast  into a nice form found experimentally by Borland and
Dennis
\begin{equation*}\label{8-4}
|x_1|+|x_2|+|x_3|+|x_4|+|x_5|+|x_6|+|x_7|\leq 4,
\end{equation*} where
\begin{eqnarray}\label{3}
x_1=\lambda_1+\lambda_2+\lambda_3+\lambda_4-\lambda_5-\lambda_6-\lambda_7-\lambda_8\nonumber\\
x_2=\lambda_1+\lambda_2+\lambda_5+\lambda_6-\lambda_3-\lambda_4-\lambda_7-\lambda_8\nonumber\\
x_3=\lambda_1+\lambda_3+\lambda_5+\lambda_7-\lambda_2-\lambda_4-\lambda_6-\lambda_8\nonumber\\
x_4=\lambda_1+\lambda_4+\lambda_6+\lambda_7-\lambda_2-\lambda_3-\lambda_5-\lambda_8\nonumber\\
x_5=\lambda_2+\lambda_3+\lambda_6+\lambda_7-\lambda_1-\lambda_4-\lambda_5-\lambda_8\nonumber\\
x_6=\lambda_2+\lambda_4+\lambda_5+\lambda_7-\lambda_1-\lambda_3-\lambda_6-\lambda_8\nonumber\\
x_7=\lambda_3+\lambda_4+\lambda_5+\lambda_6-\lambda_1-\lambda_2-\lambda_7-\lambda_8\nonumber
\end{eqnarray}
Borland and Dennis numerical data were inconclusive for the
system $\wedge^3\mathcal{H}_8$  described by 31 inequalities.
One may wonder whether they
can be written in a nice symmetric form.
\end{rem}

\section*{References}
\begin{harvard}


\item[] Bernstein I, Gelfand I and Gelfand S 1973
 {\it Russian Math. Survey} \textbf{28}(3) 1--26

\item[] Borland R E and Dennis K 1972
{\it J. Phys. B: At. Mol. Phys.} \textbf{5} 7--15

\item[] Bravyi S 2004 {\it Quantum Inf. Comp.} \textbf{4} 12

\item[] Christandl M and Mitchison G 2004
{\it Preprint\,} quant-ph/0409016

\item[] Christandl M, Harrow A W and Mitchison G 2005
{\it Preprint\,} quant-ph/0511029

\item[] Clauser J F, Horn M A, Shimony A and Holt R A 1969
{\it Phys. Rev. Lett.} {\bf 23} 880

\item[] Coleman A J 1963 {\it Rev. Mod. Phys.} \textbf{35}(2) 668

\item[] Coleman A J 2001 {\it Int. J. Quant. Chem.} \textbf{85}
196--203

\item[] Coleman A J and Yukalov V I 2000
{\it Reduced density matrices: {C}oulson's challenge} (New York:
Springer)

\item[] Coulson C A 1960 {\it Rev. Mod. Phys.} \textbf{32}(1960)

\item[] Franz M 2002
{\it J. Lie Theory} {\bf12} 539--549

\item[]
Gale D 1957
{\it Pacific J. Math.} \textbf{7} 1073--1082

\item[] Han Y-J ,  Zhang Y-Sh and  Guo G-C 2004
{\it Preprint\,} quant-ph/0403151

\item[] Higuchi A, Sudbery A  and Szulc J 2003 {\it Phys. Rev. Lett.} \textbf{90}
107902

\item[] Higuchi A 2003
{\it Preprint\,}  quant-ph/0309186

 \item[] Klyachko A 2004
{\it Preprint\,} quant-ph/0409113

\item[] Macdonald I G 1991
 {\it London Math. Soc. Lecture Notes}
 \textbf{166} 73--99

\item[] Macdonald I G 1995
 {\it Symmetric functions and {Hall} polynomials}
(Oxford: Clarendon Press)

\item[] M\"uller C W 1999
{\it J. Phys. A: Math. Gen.} \textbf{32} 4139--48

\item[] Pitowsky I 1989 \emph{{Quantum} {Probabiliy} -- {Quantum} {Logic}}
(Berlin: Springer)

\item[]
Pitowsky I and Svozil K 2001
{\it Phys.  Rev. A} \textbf{64}  014102

\item[]
Schr\"odinger E 1935
{Proc. Amer. Phil.  Soc.} \textbf{124} 323--338.

\item[] Stillinger F H \etal 1995 {\it Mathematical challenges
from theoretical/computational chemistry} (Washington: National
Academy Press)

\item[] Vinberg E and Popov V 1992 {\it Invariant theory} (Berlin:
Springer)

\end{harvard}

\end{document}